# Biological Consequences of Tightly Bent DNA: The Other Life of a Macromolecular Celebrity

Hernan G. Garcia[1], Paul Grayson[1], Lin Han[2], Mandar Inamdar[2], Jané Kondev[3], Philip C. Nelson[4], Rob Phillips[2], Jonathan Widom[5], Paul A. Wiggins[6]

[1] Department of Physics, California Institute of Technology, Pasadena CA 91125, USA
[2] Division of Engineering and Applied Science, California Institute of Technology, Pasadena CA 91125, USA
[3] Department of Physics, Brandeis University Waltham, MA 02454, USA
[4] Department of Physics and Astronomy, University of Pennsylvania, Philadelphia PA 19104, USA
[5] Department of Biochemistry, Molecular Biology, and Cell Biology, Northwestern University, Evanston IL 60208, USA
[6] Whitehead Institute, Cambridge MA 02142, USA

October 21, 2006

## Abstract

The mechanical properties of DNA play a critical role in many biological functions. For example, DNA packing in viruses involves confining the viral genome in a volume (the viral capsid) with dimensions that are comparable to the DNA persistence length. Similarly, eukaryotic DNA is packed in DNA-protein complexes (nucleosomes) in which DNA is tightly bent around protein spools. DNA is also tightly bent by many proteins that regulate transcription, resulting in a variation in gene expression that is amenable to quantitative analysis. In these cases, DNA loops are formed with lengths that are comparable to or smaller than the DNA persistence length. The aim of this review is to describe the physical forces associated with tightly bent DNA in all of these settings and to explore the biological consequences of such bending, as increasingly accessible by single-molecule techniques.

## 1. Tightly Bent DNA is a Fact of Life

In a decade whose most notable scientific achievement was the sequencing of the human genome, most discussions of DNA center on its information content. On the other hand, many of the mechanisms by which genetic information is stored and used involve deforming the DNA. Indeed, tightly bent DNA is a fact of life with biological consequences. Figure 1 shows three distinct examples of the way in which genomic DNA is subjected to tight bending. The aim of this review is to consider the physical cost and biological consequences of these different examples of DNA bending.

The problems we consider can be divided into two broad classes that involve tightly bent DNA: i) genomic packing, ii) transcriptional regulation. Often, genomic packing involves bending DNA on scales that are small in comparison with the persistence length, which is the length scale on which DNA is typically bent by thermal fluctuations. Similarly, transcriptional regulatory architectures often involve the formation of DNA loops.

The persistence length of a polymer is defined as

$$\xi_{\mathrm{p}} = \kappa / k_{\mathrm{B}} T, \qquad (1)$$

where $\kappa$ is the flexural rigidity of the filament[1], and $k_{\mathrm{B}}T$ is the thermal energy scale, around 4 pN nm (or 0.6 kcal/mole). The idea of the persistence length is that it defines the scale over which a polymer remains roughly unbent in solution. At longer scales, thermal fluctuations result in spontaneous bending of the DNA. For DNA, the persistence length has a value of ~ 50 nm (~ 150 bp). Scales larger than the persistence length are typical of those that DNA assumes in most *in vitro* molecular biology experiments such as single-molecule DNA pulling experiments.[2,3] DNA bending has been exhaustively studied in this regime. When DNA is bent on a scale shorter than $\xi_{\mathrm{p}}$, we refer to it as tightly bent, implying that the energy cost to effect such bending is large compared to $k_{\mathrm{B}}T$. Interestingly, in many of the most important biological processes, DNA adopts tightly bent configurations.

A review of these topics is timely since work over the last decade has illustrated the way in which the mechanical properties of DNA can be used as a tunable dial to elicit particular biological responses. For example, precise control of the level of gene expression can be achieved by small changes in the genomic positions of transcription factor binding sites that induce DNA looping. Similarly, the role of forces in the viral life cycle can be explored in DNA packing and ejection experiments by using DNA length as a tunable dial. One of the intriguing outcomes of this line of thought is that problems that appear only distantly related when viewed strictly from the biological perspective, bring precisely the same issues into focus when viewed from a physical perspective.

The outline of the article is as follows. In the first section, we examine how tightly bent DNA plays a role in the lifestyle of bacterial viruses (bacteriophage). As a result of recent measurements of the forces that build up during DNA packing, there has been a surge of interest in the energetics of DNA packing and ejection. The second section describes another example of how genomic packing requires tightly bent DNA, but highlighting the role of bending of nucleosomal DNA in eukaryotes. The final section explores the connection between DNA mechanics and gene expression in systems that exploit DNA looping as part of their regulatory architecture. This section focuses on the difficulties in reconciling the *in vitro* and *in vivo* pictures of DNA mechanics. Space does not permit an in-depth discussion of the intriguing question of *how* DNA mechanics is compatible with tight bending, that is, how DNA artfully contrives to appear stiff at scales comparable to the persistence length and yet adopts a variety of tightly bent configurations in the presence of proteins as shown in Figure 1.[4]

Though we concentrate on three case studies that are at the center of our own research efforts (bacteriophage DNA packing, eukaryotic DNA packing, DNA looping in bacterial transcriptional regulation), tightly bent DNA is much more widespread.[5] In that sense, this is a review of *ideas* on tightly bent DNA as illustrated by particular case studies, not a complete survey of the wide variety of different biological examples.

# 2. DNA Viruses

Many double-stranded DNA viruses have a capsid (the protein shell containing the genome) with typical dimensions of 30 to 100 nm. This capsid houses the entire viral genome, which is packaged during viral assembly. Since the genome typically has a length in excess of 10 microns, it must be tightly bent to fit into such a small protein capsid. The physical processes of genome packing and ejection in viruses raise a variety of interesting questions. How tightly is DNA bent within a virus, and what effect does this have on its lifecycle? How does DNA move from its tightly bent state within a capsid to being free within the cytoplasm of the infected cell?

## *1. The Structure of Viral DNA*

In this section, we will focus on viruses that enclose DNA within icosahedral capsids, such as herpes simplex virus 1 (HSV-1) and bacteriophage λ, as well as nearly-icosahedral asymmetric capsids (such as T7). To get a sense of the degree of confinement, it is useful to compare the capsid dimensions, 30 to 100 nm[6], to the persistence length, $\xi_P \approx 50$ nm. That is, the radius of the capsids is generally less than $\xi_P$. This means that even the outermost of the many loops of DNA within the capsid is bent at a radius smaller than $\xi_P$. Such a highly curved structure is unlikely for free DNA; even a loosely packed eukaryotic virus such as HPV-1 (diameter ≈ 60 nm[7,8]) contains its DNA in a volume thousands of times smaller than the space it would occupy if allowed to diffuse freely in solution.

Another measure of the DNA compaction is given by comparing the volume of the DNA to the volume of the capsid. For example, the length of the bacteriophage λ genome is 16 μm, and it is stored in a 58 nm diameter spherical capsid.[9] Taking DNA to be a cylinder 2 nm in diameter, the λ genome takes up a volume of roughly 50,000 $nm^3$ which should be compared to 100,000 $nm^3$, the approximate volume available within the capsid. This corresponds to a solution DNA density of about 500 mg/mL.

Early X-ray scattering experiments showed that DNA within bacteriophages is tightly packed into a nearly crystalline hexagonal array, forming the basis for models of the arrangement of DNA within the capsids.[9] These were followed by cryo-electron microscopy measurements that used averaging of tens of images to reveal a picture of the many concentric rings of DNA within bacteriophage capsids.[10] The clearest pictures of tightly-bent DNA in viruses come from recent asymmetric cryo-electron microscopy reconstructions. An example is shown in Figure 2. These studies combine data from thousands of particles to produce three-dimensional images of the capsid and genomic DNA, allowing the visualization of several layers of DNA loops within the capsid.[11-13]

## *2. Models of Tightly-Packed Viral DNA*

To gain intuition about the forces involved in DNA packing and ejection from viruses, many models of tightly bent DNA have been proposed.[14-20] The force due to bending is small during the initial stages of packing. However, as more DNA is forced into the capsid, the DNA takes up increasing amounts of available space and loops must be produced at smaller radii, increasing the force. The resulting DNA structure, thought to involve concentric loops of DNA arranged at decreasing radii about a single axis, is referred to as an "inverse spool". Alternative models of the packed DNA structure have also been proposed[21], but the asymmetric cryo-electron microscopy structures described above strongly support the inverse spool model. In one of the original models of the energetics of viral DNA packing[14], the DNA is assumed to be packed tightly into an inverse spool, with strands touching each other so that they are locally aligned on a square lattice, with an interstrand separation $d = 2$ nm. Applied to bacteriophage $\lambda$, this model predicts that the DNA loops in the center of the capsid have a radius as small as ~ 10 nm.

Due to their high negative charge, neighboring DNA loops do not touch each other, but are pushed apart by electrostatic and hydration forces.[22-27] The radius of the innermost loop will therefore be determined by an equilibrium between bending forces and the DNA-DNA interactions. This effect was taken into account in subsequent models of DNA packing.[16-18,28-30] These models are generally consistent with each other, but they focus on different kinds of predictions, such as the structure of the DNA, the forces and pressures involved in DNA confinement, and the effect of ions and DNA condensing agents. A recent advance is the construction of a model where all parameters were matched to conditions of an experiment on bacteriophage λ; the predictions could then be compared directly to experimental results without fitting, giving weight to the correctness of the model. According to this model, the inner loop will be at the extremely tight radius of ~ 3 nm.[31]

To further explore the predictions of these models, we can make a simple estimate of how much force is required to bend DNA to various amounts during the packaging of a bacteriophage λ capsid, which has a radius of about 29 nm. The energy to bend a DNA segment of length ΔL into an arc of circle of radius $R$ is given by[1]

$$\Delta E_{\text{bend}} = \Delta L \cdot \frac{1}{2}\kappa R^{-2} = \Delta L \cdot \frac{1}{2} k_{\text{B}} T \xi_{\text{p}} R^{-2}. \tag{2}$$

This implies that inserting a segment of DNA of length $\Delta L$ into the capsid, when it must be bent at this radius, will require a force of

$$F_{\text{bend}} = \frac{1}{2} k_B T \xi_{\text{p}} R^{-2}. \tag{3}$$

If we use a radius $R = 29$ nm this results in a force of order 0.12 pN, a relatively small force compared to the maximum forces exerted by molecular motors which are typically in the pN regime. The force required to bend the DNA increases as the radius of the bent DNA decreases. At $R = 3$ nm, a force of 12 pN is required. This is a high force that implies that a strong molecular motor is required simply to overcome the bending stiffness of DNA. The required force is supplied by packing motors, which consume ~ 1 ATP/2 bp and produce forces as high as 60 pN.[32,33]

It is important to keep in mind that the energy of compressed DNA within the capsid is stored in both bending and DNA-DNA interaction components: due to the force balance, the total predicted force is exactly twice what we calculated above for DNA bending alone. However, the outer strands of DNA are bent less severely than the inner strands, so that the total energy is stored primarily in the DNA-DNA interaction. In fact, in the absence of an energetic cost to DNA bending, the DNA-DNA repulsive interactions would expand the DNA crystal in the capsid, leading to large curvatures toward the capsid center. High forces would still be produced during packing due to the DNA-DNA interactions. It is the DNA-DNA interactions that are responsible for the extremely tight bending thought to exist at the center of a phage capsid. Many authors speculate that the DNA may actually be bent so tightly that it forms 180º kinks.[9,21]

The mathematical models described above are complemented by computer simulations that aim to show how the DNA arranges itself into a spool during packing[34-38] or how it moves out of the capsid during ejection.[39,40] Simulations present plausible scenarios for the details of packing and ejection, but their most interesting features are tied to assumptions that may or may not be correct in biological situations. For example, in the work by Spakowitz and Wang[38], the arrangement of the DNA depends highly on whether it is twisted during packing. Zandi *et at.*[39] consider that the forces that pull DNA out from the capsid depend on the range and strength of attraction of DNA-binding particles (such as RNA polymerase) in the cytoplasm. This issue is elaborated on in a later theoretical study.[41] In general, these kinds of assumptions present excellent targets for experimentalists desiring to improve our understanding of DNA mechanics in bacteriophages.

## *3. Measurements of the packing and ejection processes*

The forces and dynamics of bacteriophage packing and ejection are being studied with a variety of innovative experimental techniques. Since a single bacteriophage possesses a complex structure and follows a complex life cycle, averaging experimental data over particles will destroy critical information. A conceptually simple but experimentally demanding solution is to study single virus particles with microscopy and pN-scale force-probe technology, revealing information without averaging. A key experiment demonstrating the power of this approach is a study in which one end of the DNA of bacteriophage ϕ29 was held in an optical tweezer during genome packing.[32] The optical tweezer experiment can be run without feedback, in which case the force generated by the packing motor reaches an equilibrium with the force applied by the tweezer, or with constant-force feedback, in which case there is a constant tension on the DNA during packing. What was seen in the no-feedback case is that the motor can produce a force around 57 pN before it stalls, making reverse slips more and more often as it approaches the stall point. Using the constant-force case, it was determined that an opposing force was building up in the capsid throughout packing, reaching a value of about 50 pN near 100% packing.

For studying the ejection process, corresponding single-molecule techniques are not practical, since it is difficult to *push* on a long piece of flexible DNA. However, osmotic pressure can be used to push on the DNA, freezing it in an equilibrium configuration where only a fraction of the DNA, from 0 to 100%, has been ejected. Though single particles are not observed with this technique, the osmotic suppression of ejection allows us effectively to take a snapshot of a single moment in the ejection process. A series of such experiments was done on λ phage, demonstrating forces as high as 10 pN (the force corresponding to 25 atm of external osmotic pressure).[31,42-44] Since ϕ29 and λ are both packed to a similar DNA density, it is unclear whether the sixfold difference in forces is caused by a difference between the phages or a difference in the experimental conditions. The experiments on ϕ29 and λ are all consistent with the models described above.

The dynamics of ejection, which is not accessible with osmotic techniques, has been most completely addressed in recent *in vivo* studies on phage T7 and ϕ29, where it was shown that DNA enters the cell over a period of 10 to 30 min.[45,46] In this case, the study reveals the extent to which the force built up by DNA can drive the ejection. For ϕ29, it appears that force from within the capsid only drives the first part of ejection, after which an unknown cytoplasmic source of energy pulls the rest of the DNA into the cell. In the case of T7, force within the capsid does not have any apparent effect on the ejection process, and the entire DNA strand is translocated at a constant, relatively slow speed by RNA polymerase.[47]

The λ genome is known to completely enter the cell in less than 2 min, according to cyclization times and the *dam*-nuclease assay.[48] However, no lower bound exists for this transfer time; we do not know how fast the λ genome can unwind from its spool. Quantitative data about λ ejection, combined with the equilibrium force measurements, could confirm or invalidate models of the DNA ejection process.

One interesting related experiment addressed the issue using lipid vesicles containing LamB (the receptor to which phage λ attaches and which induces ejection) and filled with ethidium bromide.[49] When the DNA was ejected from λ particles into the vesicles the ethidium bromide binds to the entering DNA, causing an increase in fluorescence. The timescale of ejection as determined by this experiment was approximately 30 s. However, only ~1000 molecules of ethidium bromide were present in each of the vesicles, so that the experiment was only capable of measuring the first few kbp of DNA entry. The vesicles themselves were approximately 100 nm in diameter, so that the DNA was entering a region where it would continue to be highly bent. It is also important to realize that this experiment measures the bulk fluorescence of the entire phage ejection reaction, rather than the fluorescence of individual phage genomes, so the observed fluorescent signal will be a combination (mathematically, a convolution) of the initiation process and the actual genome transfer. Recent single molecule experiments designed to address all of these issues show that the genome transfer is actually much faster than initiation, with a timescale of about 10 s.[50]

Figure 3 shows a beautiful experiment which illustrates phage that have ejected their genomes into a lipid bilayer vesicle in a way that is analogous to the experiment on ejection dynamics. One of the most interesting features of the ejected DNA which also bears on the issue of charge interactions is that the DNA within the vesicle is collapsed into a toroid. More generally, these *in vitro* experiments on DNA toroids may help shed light on the physical forces associated with tightly bent DNA.[51,52]

For the first time, recent *in vitro* studies of T5 have described the dynamics of DNA ejection at both the bulk[54,55] and single particle[56] levels. In T5, it appears that nicks present in the genome cause the ejection to halt temporarily at defined locations. The ejection proceeds between these halting points extremely quickly, within one frame of video: it is now clear that DNA can eject at a rate of at least 75 kbp/s, but again, no lower bound can be placed on the transfer time.

## 4. Future work on DNA bending in viruses

We have seen that DNA is tightly bent within many viruses; in the bacteriophages in particular, it may be bent nearly to the limit of DNA flexibility, with a radius of curvature of roughly 3 nm. A handful of experiments has been done to investigate how the DNA unpacks, and it appears that different phages follow very different ejection mechanisms, with some ejecting tens of kbp in a fraction of a second and others taking minutes to release their genomes. However, each phage has been studied with different techniques, so it is hard to make cross-species comparisons, and we do not yet have a complete picture of the packing and ejection process for any phage. The versatility of bacteriophage λ suggests that a complete set of studies may soon be possible, using all of the *in vivo* and *in vitro* techniques described above. It will be particularly interesting to learn how the DNA is wound into the capsid, what parts spin or twist during packing and ejection, and what kind of frictional forces result from the motion of the DNA.

Bacteriophages have long served as model systems for understanding a variety of processes in biology; by studying DNA bending in phages, we gain insight into the operation of similarly-constructed eukaryotic viruses as well as DNA packing and transport in general.

# 3. DNA Packing in Eukaryotes

Like viruses, eukaryotic cells pack their genomes by tightly bending them. In these cells, chromosomal DNA is packed in a hierarchical structure. At the lowest level in the hierarchy (and our prime concern here), DNA is wrapped in 147 base pair segments roughly $1\frac{3}{4}$ times around a protein complex (the histone octamer) to form a structure known as the nucleosome as shown in Figure 4.

The nucleosomal packing motif is reiterated at short intervals along the entire length of the genomic DNA, with nucleosomes separated by short ≈ 10-50 bp-long stretches of unwrapped linker DNA. Thus, 75–90% of eukaryotic genomic DNA is wrapped in nucleosomes. The nucleosome structure itself has several particularly striking aspects. First, the DNA is exceptionally tightly bent compared to the intrinsic length scale over which DNA is flexible: the 147 bp DNA length corresponds to one persistence length, which is wrapped into loops of only ≈ 80 bp per superhelical turn. Second, the two adjacent gyres of wrapped DNA are packed extremely close together, and with their backbones in close apposition, suggesting the likelihood of strong electrostatic interactions between the two DNA gyres. Third, most of the surface of the wrapped DNA is occluded from interaction with other proteins: it is occluded on one face by close contact with the protein surface, and on the side by the close proximity of the second superhelical turn of the wrapped DNA. Since most of the genomic DNA is wrapped in nucleosomes, the preferred locations of nucleosomes may strongly impact the DNA accessibility and function of critical DNA regions.

The aim of the discussion here is to explore the consequences of the fact that the nucleosomal DNA is tightly bent on the scale of the persistence length. Despite the energetic cost of bending the DNA on these small scales, the favorable contacts between positively-charged residues on the histones and the negative charges on the DNA suffice to overcome this energetic penalty.[57,58] Indeed, in some sense the question is not why do nucleosomes form, but rather, how do proteins that need to gain access to nucleosomal DNA do so?

## 1. Equilibrium and Dynamics of Nucleosomal DNA Accessibility

X-ray crystal structures of nucleosomes[59,60] (see Figure 4) show the wrapped DNA to be largely inaccessible to the many protein complexes that must bind it for essential DNA transactions such as replication, transcription, recombination, and repair.[61-65] However, as is often the case in biology, the structure appears to be tuned for marginal net stability, with the attractive interactions slightly exceeding the elastic cost of wrapping tightly around the positively-charged protein spools. Probabilities depend exponentially on the energetics, so the probabilities overwhelmingly favor the wrapped state; but because the energetics are marginal, there will nevertheless be frequent (if short lived) unwrapping events.[66]

To see how the relevant energies compete with each other, we resort to some simple estimates. The energy associated with bending the DNA can be estimated simply by invoking a version of Equation 2 applicable to circular loops of radius $R$ and given by

$$E_{\mathrm{loop}} = \frac{\pi \xi_{\mathrm{p}} k_{\mathrm{B}} T}{R}, \qquad (4)$$

where $\xi_{\mathrm{p}}$ is the DNA persistence length. Here we use a flexural rigidity, $\kappa = \xi_{\mathrm{p}} k_{\mathrm{B}} T$. To get an estimate of the energy scale, we note that the radius of curvature at the centerline of the DNA is roughly 4.5 nm, corresponding to an energy $E_{\mathrm{loop}} = 35\ k_{\mathrm{B}} T$. As noted above, the second key contribution to the energy comes from the interactions between the charges on the histones (positive charges) and the DNA (negative charges). Over the $1\frac{3}{4}$ times that the DNA wraps around the histone octamer, there are 14 distinct contacts each of which has a contact energy of roughly -6 $k_{\mathrm{B}} T$. These contacts between the protein core of the nucleosome and the wrapped DNA occur in patches, every DNA helical turn, when the minor groove (DNA backbone) wraps around and faces inward toward the protein core.[59,60] The contact energy can also be modeled as a continuous adhesion energy $E_{\mathrm{contact}} = \gamma_{\mathrm{adh}} L$, where $\gamma_{\mathrm{adh}}$ is an energy/length with a value of roughly $\gamma_{\mathrm{adh}} \approx -2.0 k_{\mathrm{B}} T/\mathrm{nm}$, with the minus sign signaling that this is a favorable contact. These values can be obtained by fitting this simple model to measurements on the equilibrium accessibility of nucleosomes.[67]

One of the principal puzzles posed by the function of nucleosomes is how these structures are at once stable and yet accessible to DNA-binding proteins. Restriction enzymes experience the same accessibility obstacles for action on nucleosomal DNA as do eukaryotic protein complexes, and have been used to probe the equilibrium accessibility of the wrapped DNA.[67] The basic idea behind these experiments is to measure the probability of restriction digestion as a function of burial depth of the restriction site of interest within the nucleosome. These studies reveal that stretches of the nucleosomal DNA located a short distance inside the nucleosome from one end act as though they are (unwrapped) naked DNA molecules a surprisingly large fraction (several percent) of the time, i.e., there is an equilibrium constant for dynamic unwrapping of the ends of the nucleosomal DNA on the order of 0.01 to 0.1. This equilibrium accessibility drops progressively with distance further inside the nucleosome, decreasing to $10^{-4} - 10^{-5}$ for sites located near the middle of the nucleosome.

These findings can be understood using the simple model described above based on the structure of the nucleosome. For simplicity, we assume that each contact patch contributes an equivalent net favorable free energy for DNA wrapping, and that access to sites further inside the nucleosome is achieved by starting at one end of the nucleosome, and unwrapping the DNA one helical turn at a time, breaking contact patches in succession, until enough DNA is unwrapped such that a given site is now accessible. Each broken contact costs a certain amount of net free energy, so access to sites further inside the nucleosome comes with a stepwise increasing cost in free energy, and a corresponding stepwise decrease in probability or equilibrium constant. In the continuum model described above, one assumes that the free energy cost is a continuous function of the degree of unwrapping. In particular, if the nucleosomal DNA is peeled off by an amount $x_{\mathrm{e}}$, then the free energy of the bound DNA is $F(x_{\mathrm{e}}) = (\gamma_{\mathrm{bend}} - \gamma_{\mathrm{adh}})(L - x)$. Using this simple model of the energetics of nucleosomal DNA, the configurational equilibrium constant can be computed as

$$K_{\mathrm{conf}}(x_{\mathrm{re}}) = \frac{e^{\frac{1}{k_{\mathrm{B}}T}(\gamma_{\mathrm{adh}} - \gamma_{\mathrm{bend}})(L - x_{\mathrm{re}})} - 1}{e^{\frac{1}{k_{\mathrm{B}}T}(\gamma_{\mathrm{adh}} - \gamma_{\mathrm{bend}})L} - 1} \tag{5}$$

where $x_{\mathrm{re}}$ is the depth of the site of interest, $L$ is the total length of wrapped DNA and $\gamma_{\mathrm{bend}}$ is the bending energy per unit length associated with the wrapped DNA.

The experiments described above show nucleosomal DNA to be dynamically accessible, but leave open the question of the actual rates. Two new experiments show that nucleosomes spontaneously open to allow access to at least the first 20 to 30 bp on timescales as short as 250 msec.[66] The dynamic accessibility implies that transcription factor binding sites, promoters, etc. that are buried in a nucleosome can remain active although at a significantly lower level than identical sequences that are unbound. A large number of nucleosome remodeling factors have been identified suggesting that cells may further increase the accessibility of buried sites by active mechanisms.[68-71]

Nucleosomes look like those imaged by X-ray crystallography for very short periods of time before spontaneously undergoing large scale opening conformational changes. However, these open states do not last long, typically only 10 to 50 msec, before the DNA spontaneously re-wraps. A mean-first-passage-time calculation based on a continuum version of the nucleosome-DNA adhesion picture described above allows for parameter-free predictions of the opening and re-wrapping rates as a function of distance.[72] A different experiment probing unwrapping to sites further inside the nucleosome appears to concur at least qualitatively with these predictions, showing that the unwrapping to greater depth does occur on a significantly slower timescale.[73]

## *2. Sequence-Dependence of Nucleosome Formation and Accessibility*

In our discussion of DNA packing in viruses, we showed how the length of the DNA molecule could be used as a tunable dial to alter the mechanical forces associated with the packaged DNA. DNA sequence is yet another way in which the energetics of tightly-bent DNA can be tuned and altered. The key point is that different sequences have different intrinsic *bendability*, and hence a quantitatively different tendency to form nucleosomes. In particular, the tight bending of DNA in nucleosomes causes them to prefer certain particular DNA sequences over others. DNA sequences exhibiting a greater than 5,000-fold range of affinities for wrapping into nucleosomes are documented;[74,75]

moreover, the range of affinities may be even greater, as the experiments used to measure the relative affinities may artificially underestimate the true range.

These sequence preferences are not due to particular favorable base-specific interactions[59], as would normally be the case for site-specific protein-DNA complexes. Rather, sequence-dependent nucleosome positioning represents an extreme case of indirect readout.[76] In indirect readout, sequence preferences arise from the differing abilities of differing DNA sequences to adopt particular idiosyncratic conformations required by the proteins - which for the case of the nucleosome, is dominated by the extremely tight DNA bending required.

The most important DNA sequence motifs that confer high affinity binding to the nucleosome are AA, TT or TA dinucleotide steps (that is, an A followed by another A, and so on), which recur every 10 bp, in phase with the DNA helical repeat, every time the DNA minor groove (phosphodiester backbone) rotates around to face inward toward the center of curvature of the nucleosomes protein core. A survey of high resolution X-ray crystallographic structures of DNA[5] suggests that no dinucleotide steps favor such bending into the minor groove, but, evidently, these particular steps minimize the unfavorable energetic cost. There exist also weaker preferences for certain other steps, most notably GC (that is, G followed by C) to occur exactly out of phase with the AA/TT/TA steps, every time the minor groove faces outward.

Several lines of reasoning and, more importantly, direct experimental tests[75,77] show that particular DNA sequences that are especially soft for bending (as opposed to intrinsically bent in the manner favored by the nucleosome) make particularly stable nucleosomes. The role of DNA bending in determining these sequence preferences is illustrated dramatically by comparing the free energy of cyclization (the cost to make a small loop in solution) with the free energy of nucleosome formation as shown in Figure 6. Nevertheless, the detailed molecular mechanics basis of all of these sequence preferences remains unknown and is an important topic for further research.

The biological significance of these nucleosomal DNA sequence preferences arises because they imply that nucleosomes are not distributed randomly along genomic DNA. Eukaryotic genomes utilize these sequence preferences together with the powerful force of steric hindrance - nucleosomes occupy space and cannot overlap - to encode an intrinsic nucleosome organization. The resulting *in vivo* distribution of nucleosome occupancies appears to facilitate many aspects of chromosome function, including transcription factor binding, transcription initiation, and even remodeling of the nucleosomes themselves.[78]

Interestingly, eubacterial genomes, which lack histones, nevertheless encode 11 bp-periodic distributions of AA/TT dinucleotides.[79] These are not attributable only to protein coding requirements[80,81] and instead suggest that prokaryotic genomes encode an intrinsic three dimensional organization of their own chromosomes different from, but analogous in some ways to, the intrinsic nucleosome organization encoded in eukaryotic genomes.

# 4. Tightly bent DNA in transcriptional regulation

Gene expression is subject to tight control and one of the most important mechanisms of regulation occurs at the level of transcription. Transcriptional regulation is carried out by a variety of DNA-binding proteins known as transcription factors. The two key case studies that led to the elucidation of the operon concept (the idea that there are genes that control other genes)[82], namely the *lac* operon and the λ switch, both involve DNA looping.[83,84] In these cases, the DNA binding proteins that mediate transcriptional control bind at two sites on the DNA simultaneously, looping the intervening DNA. Indeed, tightly bent DNA is a ubiquitous motif in both prokaryotic and eukaryotic transcriptional regulation. In Table 1 we highlight some of the best known examples of this regulatory architecture. In most cases, the relevant loops have lengths that are comparable to or smaller than the persistence length.

Given that the persistence length is the scale over which DNA is stiff, it is surprising that short loops play such an important role in transcription. The implicit assumption that leads to that surprise is that the effective *in vivo* DNA flexibility is the same as that measured extensively for bare DNA *in vitro*. However, such *in vitro* measurements generally only probe length scales much longer than those relevant to the structures in Figure 1.[4] In order to analyze the role of tightly bent DNA in transcriptional regulation we will focus on three physical mechanisms: (i) the *in vivo* bendability and twistability of DNA, (ii) the contribution from protein conformation and (iii) the presence of a whole battery of non-specific or nucleoid-associated DNA binding proteins which play an active role in determining structural and dynamical properties of the bacterial chromosome.

Though there are a host of interesting examples of transcriptional regulation that involve DNA looping, we focus almost exclusively on the dissection of the role of looping in the *lac* operon. The *lac* operon refers to the genes responsible for lactose metabolism in bacteria.[83] In particular, when faced with an absence of glucose and the presence of lactose, this operon will be "on" resulting in the production of β-galactosidase (and several other proteins as well), the enzyme responsible for the digestion of lactose. The challenge is to see how *in vivo* and *in vitro* experiments and modeling approaches can be used to tease out the mechanism and biological significance of DNA looping: why do genomes bother to loop?

We will focus on data in which the mechanical properties of DNA are used as an adjustable dial to tune a desired biological outcome during transcriptional regulation. In particular, we will address *in vivo* experiments like those performed by the Müller-Hill[101], Record[102] and Maher[103] groups where the level of repression is systematically measured as a function of the distance between two binding sites for Lac repressor (Figure 7A inset). For reviews on Lac repressor refer to Matthews and Nichols[104] and to Lewis[105]. In addition, we will examine corresponding *in vitro* measurements of the interaction between Lac repressor and its target DNA.

As shown in Table 1 there are many other interesting examples of DNA looping in transcriptional regulation. We focus on one such case study because in this case, there are a broad range of quantitative measurements that permit a careful comparison of results from both *in vivo* and *in vitro* experiments. These results may be used to form a coherent picture of looping in transcriptional regulation though current models fall short of a complete picture of these problems that leads to consistent, falsifiable experimental predictions. We view this as an opportunity to propose a set of careful quantitative and systematic experiments that will help decouple the contributions and importance of the different molecular players in this process.

## 1. In vivo DNA looping: Using cells as test tubes

The most common and straightforward way of characterizing the action of some regulatory motif on gene expression is by measuring relative changes in the activity or concentration of the regulated protein product. The classic reporter has been β-galactosidase. The concentration of this gene product is characterized by measuring its activity in lysed cells using a colorimetric assay.[106] The unequivocal signature for DNA looping since its discovery by Schleif and co-workers in the arabinose operon has been the modulation of gene expression as a function of the length of the DNA loop with a periodicity of roughly 11 bp corresponding to the effective *in vivo* helical pitch of DNA.[107-109] This type of experiment shows how quantitative, single-molecule mechanical properties can be extracted from cells by looking at changes in the protein expression profile of an entire population of cells. It is remarkable that changes in DNA such as making the molecule a single base pair longer or shorter, can result in such clear macroscopic effects in an ensemble of cells. An example of this kind of data for the *lac* operon is shown in Figure 7A. In many ways, the remainder of this review centers on understanding the many distinctive features of this curve which has hidden within it several intriguing clues and puzzles concerning DNA mechanics.

Precise and rich data like those shown in Figure 7A present a variety of theoretical challenges. Thermodynamic models of transcriptional regulation[111,112] have been used to extract the free energy of looping which is a measure of the cost of the looped configuration as a function of the distance between the operators.[102,113-115] These models use equilibrium statistical mechanics to describe the probability of transcription as it is modulated by the presence of the repressor and its partner looped DNA.

One of the biggest challenges in modeling the Lac repressor-loop-mediated repression lies in the fact that the free energy of the looped configuration is determined by a variety of factors. In addition to the free energy of DNA looping itself, it is also necessary to consider the geometry and flexibility of the looping protein[116,117] and the presence of non-specific binding proteins such as HU, IHF and H-NS in the background[103] (for a review of the role of these proteins in the organization of bacterial chromatin refer to Luijsterburg *et al.*[118]). Nevertheless, it is still meaningful as a first approximation to compare the *in vivo* looping energy extracted from these experiments to the energy of cyclization of DNA circles defined in section 3 at the same length scales, where the additional subtleties of the *in vivo* experiment are not present. Such a comparison is shown in Figure 7B.

There are at least three striking features of the *in vivo* looping energy in comparison with its *in vitro* counterpart. First, the minimum in the looping free energy at 70 bp does not coincide with the expected cyclization minimum at around three persistence lengths. Second, at 70 bp, there is an overall offset between the *in vitro* and *in vivo* values and, finally, a difference in the amplitude of the twist modulation. All of these features suggest that it is easier for DNA to adopt tightly bent configurations in the *in vivo* setting than would be expected from our intuition based on studies of DNA mechanics *in vitro*. In the remainder of this section, we review some of the available evidence that sheds light on the origin of these differences between *in vivo* DNA looping and *in vitro* DNA cyclization.

The position of the minimum in the *in vivo* looping free energy shown in Figure 7B suggests that for these tightly bent configurations, DNA has a lower effective persistence length than the canonical value of ~ 150 bp. Interestingly, proteins that are expected to be more flexible than wild type Lac repressor such as AraC[108,119-121] and Lac repressor mutants[122] present a different shape in their gene expression curves and, consequently, in their looping energies. In both cases the looping energy does not display a minimum. Rather, it keeps decreasing as the interoperator distance gets shorter. Various computational studies have addressed the issue of protein flexibility.[123-126] Even though the difference in the position of the minima can be accounted for, a smaller value of persistence length is still needed in order to fit the models to the available *in vivo* data.[126]

It can be argued that the main difference in the absolute value of the looping energy between cyclization and *in vivo* looping in transcriptional regulation can be accounted for by a difference in the definition of the standard states or zeros of free energy. For example, in the *in vitro* case, the reference state is defined as the uncyclized linear molecule in solution. On the other hand, in the more complex *in vivo* case, this reference state is not as clearly defined. In particular, in this case, even when not bound to specific operator sites, DNA is bound non-specifically[127], presumably resulting in a host of different looped states. The set of all of these different looped states defines the reference state for the *in vivo* case. Additionally, the presence of negative supercoiling inside the cell[128] and of non-specific DNA binding proteins such as the histone-like HU[103] have been shown to be factors that can modify the reference energy. Without knowledge of how this reference state is determined, no absolute comparison between *in vivo* and *in vitro* data can be made.

The third key feature calling for attention is the unexpectedly small amplitude of the periodic modulation in the *in vivo* looping free energy. One explanation for this difference between the effective *in vivo* looping free energy and the cyclization free energy could be a higher DNA twistability of tightly bent DNA.[75] So far, the available computational models have not been able to show how protein flexibility alone can account for this difference.[126] Müller-Hill and co-workers proposed that such an apparent lower modulation could be explained if different loop species were present.[134] These different species could correspond to different topoisomers[135], different orientations of the operators with respect to the symmetric binding heads or different conformations in Lac repressor[124,126,136] which are supported by *in vitro* evidence (see below).

Understanding DNA looping *in vivo* in bacteria requires understanding the role of tightly bent DNA in these systems. However, the *in vivo* approach only yields a single quantity, namely, the looping free energy. The problem is that this quantity reflects not only the mechanical properties of DNA, but also the effect of protein flexibility and the effects of other proteins bound to the DNA. Addressing this problem from the *in vitro* perspective of biochemistry allows for a more controlled characterization of the effect of the different molecular players in this process. We conclude this section by reviewing some recent and classical *in vitro* studies of DNA looping by Lac repressor.

## 2. *In vitro* DNA Looping: DNA Mechanics One Molecule at a Time

Complex cellular processes like those described above can be tackled *in vitro* using the tools of solution biochemistry and single molecule biophysics. Both of these approaches have been unleashed on the problem of DNA looping in the context of transcriptional regulation.

Bulk binding assays involving DNA-binding proteins such as Lac repressor and their DNA targets measure the affinity of these proteins for configurations with different looping lengths or degree of supercoiling, for example. Filter binding assays and electrophoretic mobility shift assays are two examples of these kinds of technique. In the gel-shift assay, the electrophoretic mobility of a given fragment of DNA is measured both in the absence and presence of the DNA-binding protein of interest. When the DNA-binding protein binds to the DNA fragment, it changes its motility in the gel and is detected as a new band. By tuning the concentration of the binding protein, as well as controlling variables dictating DNA mechanics (such as the looping length or the degree of supercoiling), it is possible to measure how these mechanical variables alter the binding probability.

In contrast to the *in vivo* observation, using the gel shift assay, Krämer *et al.* determined that the probability of looping decreases as the distance between operators on a linear DNA fragment decreases from 210 to 60 base pairs.[137] This result agrees with the observations by Hsieh *et al.* using the filter binding assay, whose quantitative results are shown in Figure 8.[129] This disagreement in the behavior of the looping free energy as the distance between operators decreases between the *in vivo* and *in vitro* experiments is a stark reminder of the challenge of reconciling the *in vitro* and *in vivo* pictures of DNA mechanics in general, and protein-mediated looping in specific.

Similar experiments have been used to characterize the role of supercoiling by using supercoiled plasmids.[130,134,138] Interestingly, these experiments reveal an increase in the affinity of the Lac repressor to a single site showing that negative supercoiling favors binding. Most importantly though, a dramatic increase in the looping probability was observed. This increase in looping probability is revealed in changes in the protein-DNA complex dissociation times that varied from 2 hours to more than 20 hours.[134] These experiments also suggested that the looping energy does not change much over distances between 100 and 500 bp for a negatively supercoiled template.[130] However, this could not be confirmed because the distance between operators was not systematically varied. A decrease in the twist modulation was also observed, suggesting, as was mentioned in the previous section, that multiple topoisomers coexist for certain separations.[134]

These results have been supplemented with several other classes of experiments, some of which involve the *direct* observation of individual loops. Using microscopy techniques such as electron microscopy[134,137] and atomic force microscopy[132], individual loops can be observed and key parameters such as the loop length can be measured. These experiments have been valuable not only in the context of Lac repressor, but also in identifying different looping motifs in complex cis-regulatory regions in eukaryotic systems.[94]

Another important class of experiments that have shed light on the mechanics of DNA looping *in vitro* are single-molecule measurements using the Tethered Particle Motion (TPM) method as shown in Figure 9.[139] TPM was first used by Finzi and Gelles in the context of DNA looping to directly detect Lac repressor mediated loop formation and breakdown, and to measure the kinetics of such processes.[140] In this method, a DNA molecule is tethered between a microscope slide and a microsphere which is large enough to be imaged with conventional optical microscopy. The Brownian motion of the bead serves as a reporter of the underlying DNA dynamics. In particular, when the molecule is unlooped, the tether has its full length and the excursions of the bead are large. When the DNA is looped, the tether is shortened and the excursions are reduced.[141-148] Thus, modulations in motion reflect conformational changes in the tethered molecule. This method has recently revealed[132,149] the presence of two-looped states which is consistent with the presence of multiple configurations observed using FRET[150,151], electron microscopy studies[152] and suggested by x-ray crystallography studies[116]. All of these experiments suggest an important role for *protein* flexibility. A more sophisticated technique which has been successfully applied to Gal repressor is the magnetic tweezer assay.[153] In this case, the tether can be stretched and twisted as the dynamics of looping and unlooping are followed leading to measurements of the underlying kinetics, thermodynamics and supercoiling dependence.

Even though Lac repressor can loop in the absence of any other DNA binding proteins, other systems such as GalR require the presence of the non-specific DNA binding protein HU.[153] HU has been proven to alter the effective flexibility of DNA.[154] However, this issue has not been studied systematically in the context of DNA looping or in the presence of other non-specific binding proteins such as H-NS and IHF.

In spite of more than two decades of investigation, there is still no comprehensive or quantitative link between *in vivo* and *in vitro* studies of looping and DNA conformation (Figures 7 and 8). For instance, although it is known qualitatively that nucleoid-associated-protein binding and supercoiling can both significantly enhance looping efficiency, we still do not know whether these mechanisms are sufficient alone or in tandem to explain the dependence of repression on inter-operator spacing observed in a host of biological systems. To get to the bottom of these questions will require further systematic and quantitative experiments. In particular, systematic experiments which vary specific experimental tuning parameters (operator distance, sequence, concentration of nucleoid-associated proteins) need to be performed.

Most of the *in vivo* data on DNA mechanics as revealed by transcriptional regulation suggests an increased DNA flexibility, signaling that there is more to the effective *in vivo* looping free energy than is offered by the wormlike chain model alone. Interestingly, recent *in vitro* experiments also suggest short-length scale anomalies in DNA mechanics[4,75,155,156], even though no consensus has been reached.[133] In order to fully understand the role of tightly bent DNA in transcriptional regulation the contribution of the different molecular players (intrinsic DNA mechanics, architectural proteins, transcription factors, supercoiling) has to be decoupled.

# 5. Conclusion

We have argued that tightly bent DNA is a common feature in living organisms. The packing of genomic DNA in viruses, prokaryotes and eukaryotes involves both indirect (confinement by protein capsids) and direct (architectural proteins such as HU and histones) interactions between DNA and proteins which lead to highly deformed DNA configurations. Similarly, transcriptional regulation in prokaryotes and eukaryotes routinely requires the formation of DNA loops involving DNA segments that are shorter than the persistence length.

Interestingly, in all of the examples described in this review, the physical mechanisms associated with tightly bent DNA lead to biological consequences. For example, because of the energetic costs associated with genome confinement, bacteriophage have extremely strong molecular motors to pack their DNA. Eukaryotic DNA is packed in nucleosomes, requiring a bevy of proteins to rearrange nucleosomes. In addition, nucleosomes preferentially bind to DNA sequences that are easy to bend. Combinatorial control in transcriptional regulation is often mediated by transcription factors that induce DNA looping. In each of these cases, there is a direct connection between the physical properties of DNA and its biological function.

These problems have been addressed by scores of researchers using a wide variety of different experimental and theoretical techniques. Interestingly, the flow of information and understanding works in two ways: fundamental studies of DNA mechanics in these various settings reveal new biology; and fundamental studies of the basic biology reveal striking new aspects of DNA mechanics. One of the surprising outcomes of work in this area has been the realization that DNA mechanics can play a significant role in dictating biological function. Further, it has become increasingly possible to dial in different DNA mechanical properties (using DNA sequence and length as tuning parameters) as a way of either controlling or exploring different biological processes.

One of the significant outstanding challenges is that our *in vitro* and *in vivo* pictures of the mechanical properties of DNA are inconsistent. These inconsistencies could only be appreciated when the problems were viewed quantitatively. The resolution of these outstanding issues will require systematic, quantitative experiments in both the *in vitro* and *in vivo* settings. As a result, there remain a wide variety of important unanswered questions concerning the mechanical behavior of tightly bent DNA and how it relates to biological function which will keep researchers from both the biological and physical sciences busy for a long time to come.

## Acknowledgments

We are extremely grateful to a number of people who have given us both guidance and amusement in thinking about these problems (and some for commenting on the manuscript): Sankar Adhya, David Bensimon, Laura Finzi, Bill Gelbart, Jeff Gelles, Uli Gerland, Shura Grosberg, Jack Johnson, Jason Kahn, John Maddocks, Jim Maher, Ian Molineux, Wilma Olson, Prashant Purohit, Michael Rubenstein, Robert Schleif, Andy Spakowitz, Elizabeth Villa and Nils Walter. HG is grateful for support from both the NSF funded NIRT and the NIH Director's Pioneer Award. JK acknowledges the support of NSF DMR-0403997 and is a Cottrell Scholar of Research Corporation. PN acknowledges NSF Grant DMR04–04674 and the NSF-funded NSEC on Molecular Function at the Nano/Bio Interface DMR04–25780. RP, PG, MI and PW acknowledge the support of the NSF and the NIH Director's Pioneer Award.

# Figures and tables

**Figure 1.** Biological examples of tightly bent DNA. (A) Transcription factor mediated DNA looping, (B) DNA packing in the nucleosome, (C) DNA packing in bacterial viruses (Courtesy of David Goodsell, Scripps Research Institute, La Jolla, California).

**Figure 2.** Images of packaged viral DNA. This figure shows two recent reconstructions using cryo electron microsopy of the packaged DNA. (A) Phage ε15 DNA from Jiang *et al.*[11]— reconstruction without symmetry. The size of the scale bar is 20 nm. (B) Phage P22 DNA, with the portal shown in red (courtesy of Gabriel Lander and Jack Johnson). This view is looking into the capsid at the portal (the entry site for DNA) and the green hoops reflect density corresponding to the packed DNA.

**Figure 3.** Images of DNA ejected into a lipid bilayer vesicle. Empty capsid are distinguishable from their full counterparts because the full capsids are much darker.[53]

**Figure 4**. Structure of the nucleosome. Two orthogonal views of the nucleosome showing the wrapping of the DNA around an octameric histone protein core.[59] The core histone proteins are colored yellow, red, blue and green for histone H2A, H2B, H3 and H4, respectively. There are two copies of each histone in the core histone octamer. The two strands of the double helix are colored cyan and brown. The diameter of the nucleosome is roughly 11 nm and its height is roughly 6 nm.

**Figure 5.** Configurational equilibrium constant. Measured values of equilibrium accessibility and corresponding results from the model of nucleosome energetics. The inset shows a schematic of the coordinate system used to define the burial depth of the binding site of interest.

**Figure 6.** Free energy of cyclization and nucleosome formation. Difference free energies for wrapping of different 94 bp DNAs around the core histone H32H42 tetramer are plotted against the difference free energies of cyclization for these same DNAs.[74] The line illustrates the least-squares fit to the data. The slope of the line is one, implying that the entirety of the difference in affinity for wrapping around histones can be explained by the difference in the ability to cyclize.

**Figure 7.** *In vivo* DNA looping by Lac repressor and the *in vitro* challenge. (A) Data from Müller *et al.*[101] showing repression as a function of distance between operators. (B) Change in looping free energy obtained from the Müller-Hill data (black) and theoretical prediction of the energy of cyclization of a DNA molecule based on the worm like chain model[110] and assuming a volume for *E. coli* of $V_{cell} \approx 1\mu m^3$ such that $\Delta F_{cyclization} = -\ln(J_{cyclization} \cdot V_{cell})$. Note that the minima in the two curves do not coincide, signaling that the *effective* looping free energy *in vivo* is not the same as the bare looping free energy deduced from *in vitro* cyclization measurements. In addition, there is an overall shift in the scales in the two cases. (Inset, B) Difference in the magnitude of the twist modulation between the looping energy obtained from the Becker *et al.*[103] data and the theoretical cyclization energy based on harmonic deformations of the base steps[75].

**Figure 8.** Effective *J*-factors for *in vitro* DNA looping. The graph is constructed by using a variety of different *in vitro* measurements to derive an *effective* looping *J*-factor, even in those cases where there was no direct measurement of *J* itself. The derived values were obtained from: i) bulk linear DNA[129], ii) bulk supercoiled DNA[130], iii) single molecule measurements[131,132], iv) DNA cyclization[75,133] and the blue curve is a theoretical curve for cyclization corresponding to an extrapolation of the elastic rod model[110].

**Figure 9.** Illustration of TPM method. Schematics of both the unlooped and looped states which show how the effective tether length is a reporter of the state of looping. Typical tethers have a length of 1000 bp and typical bead sizes are 0.2 - 1.0 μm.

**Table 1.** DNA looping in prokaryotic and eukaryotic transcriptional regulation. Loop lengths and mechanisms of action of some of the best known looping systems in bacteria and eukaryotes. Note that these loop lengths suggest tightly bent configurations since the *in vitro* measured persistence length is 150 bp.

| Molecule or locus | Mode of action | Wild type loop lengths (bp) |
|---|---|---|
| Lac repressor[83] | Repression | 92, 401 |
| AraC[85] | Repression and activation | 210 |
| Gal repressor[85] | Repression | 115 |
| Deo repressor[85] | Repression | 270, 599, 869 |
| Nag repressor[86] | Repression | 93 |
| NtrC[87] | Activation | 110~140 |
| λ repressor[84,88] | Repression and activation | ~2,400 |
| XylR[89] | Activation | ~150 |
| PapI[87,90] | Activation | ~100 |
| β-globin locus[91,92] | Activation | 40,000-60,000 |
| RXR[93] | Activation | 30-500 |
| SpGCF1[94] | Activation, domain intercommunication | 50~2,500 |
| HSTF[95] | Activation | 23 |
| p53[96] | Repression and activation | 50~3,000 |
| Sp1[97-99] | Activation | ~1,800 |
| c-Myb and C/EBP[100] | Activation | ~80 |

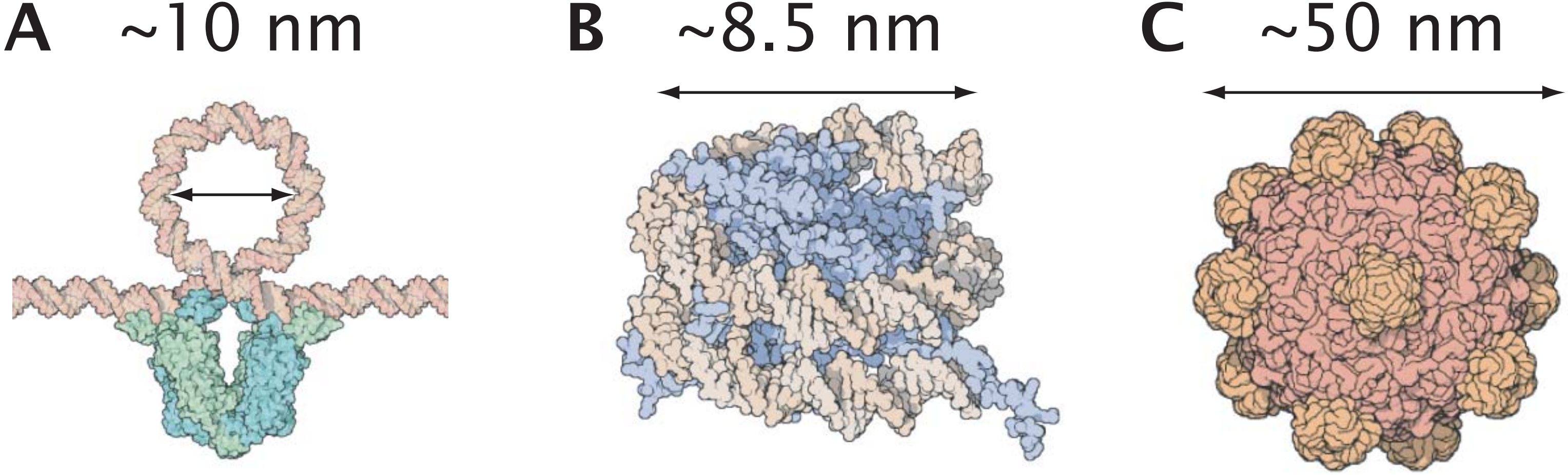

Figure 1

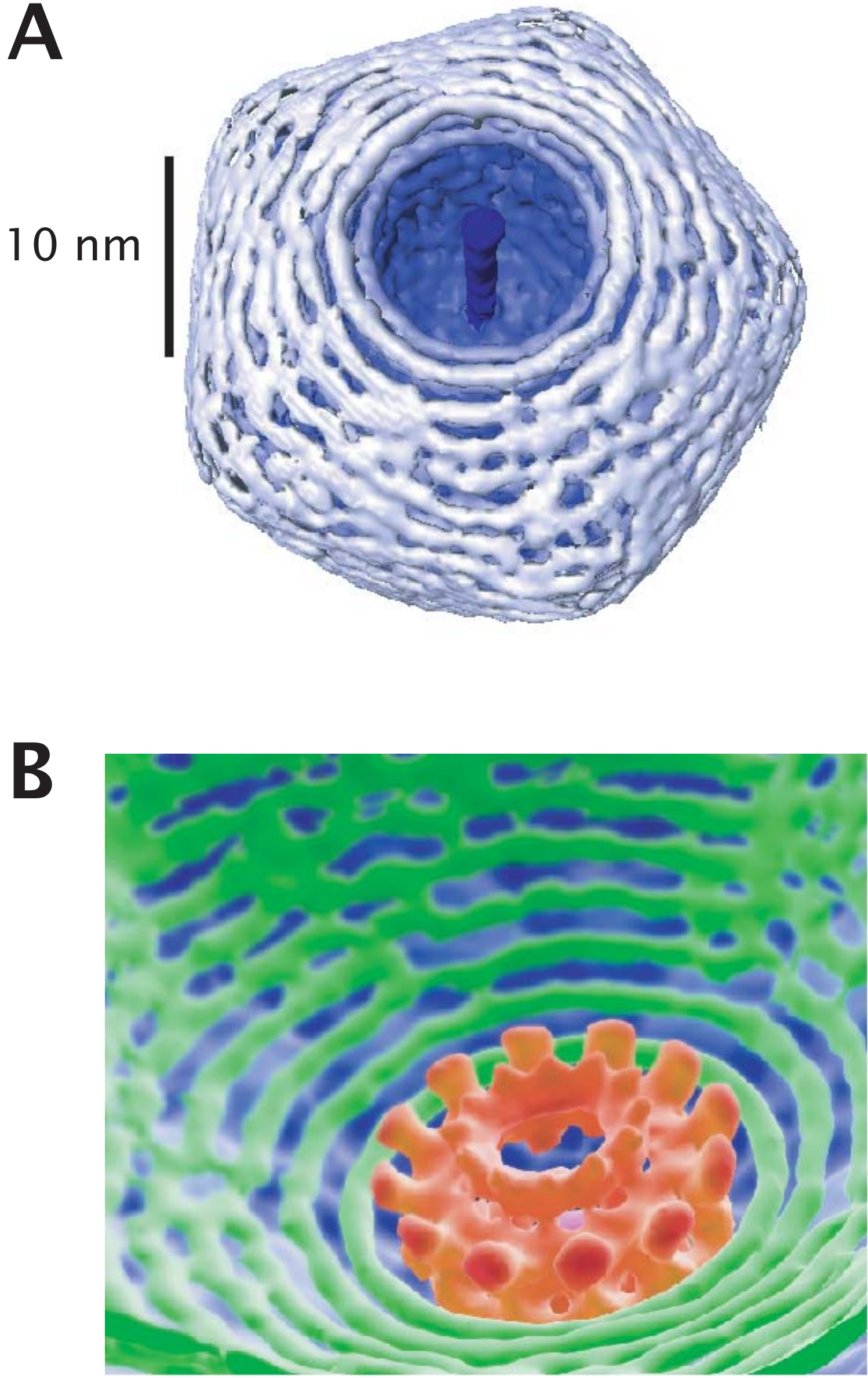

Figure 2

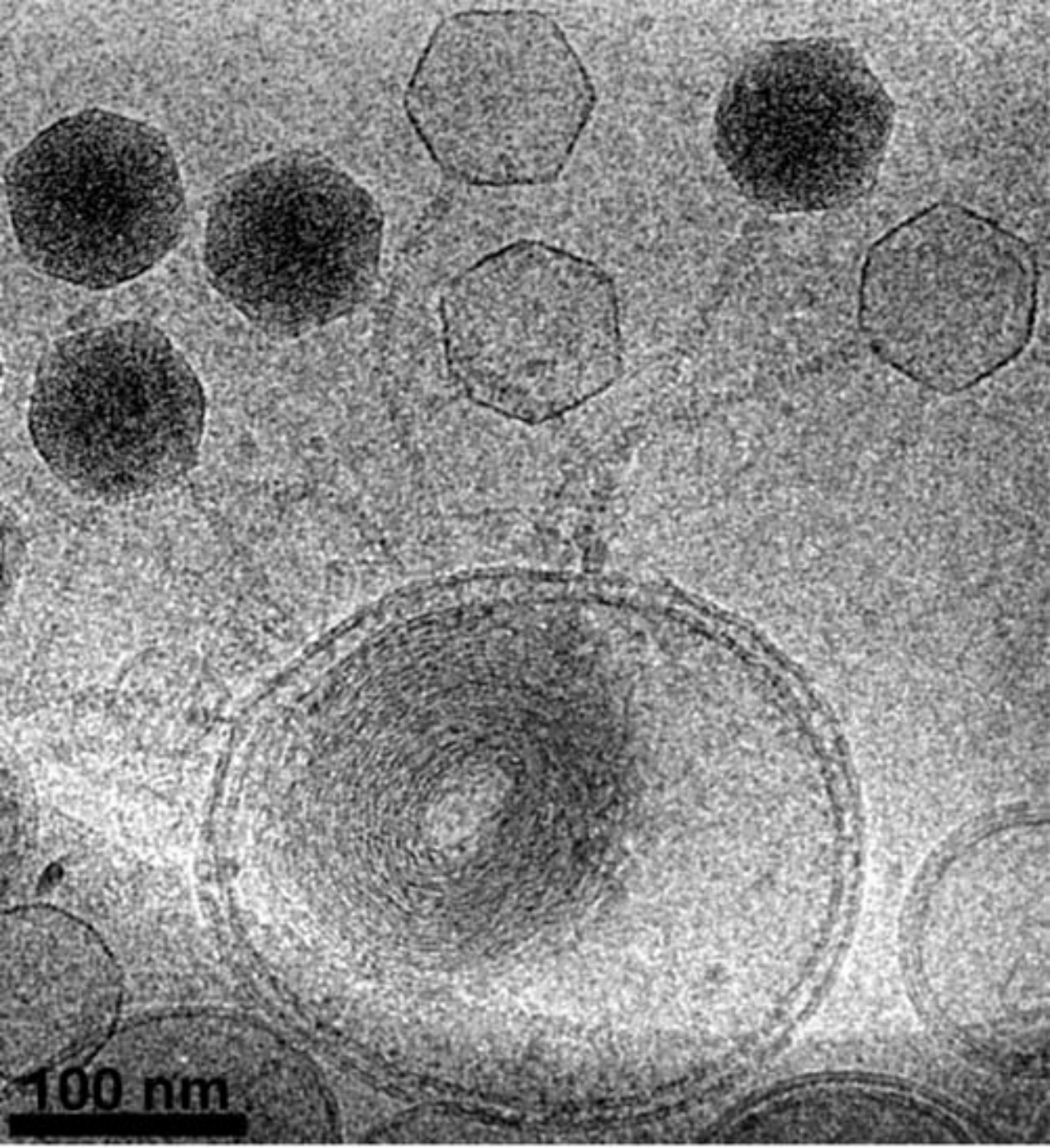

Figure 3

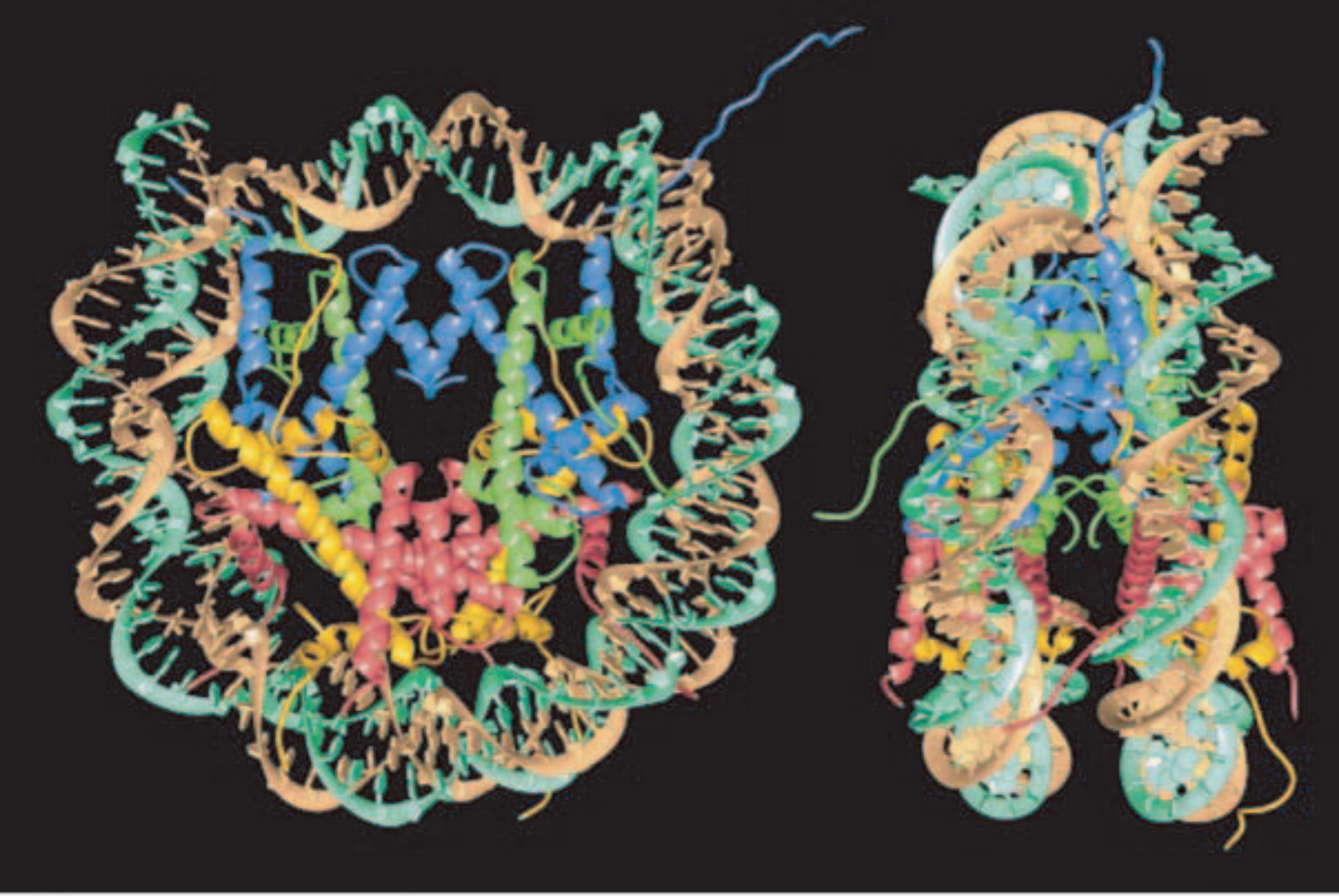

Figure 4

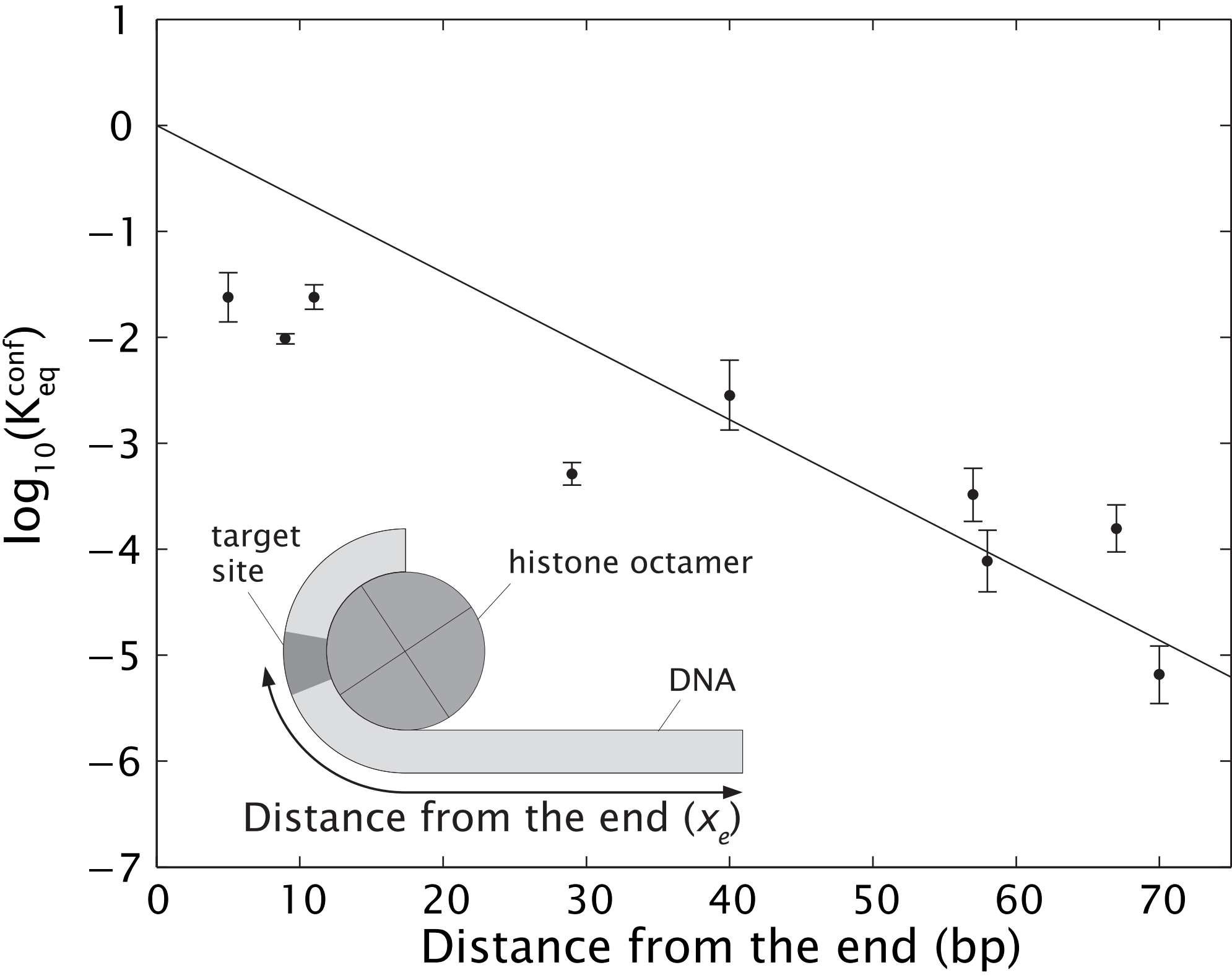

Figure 5

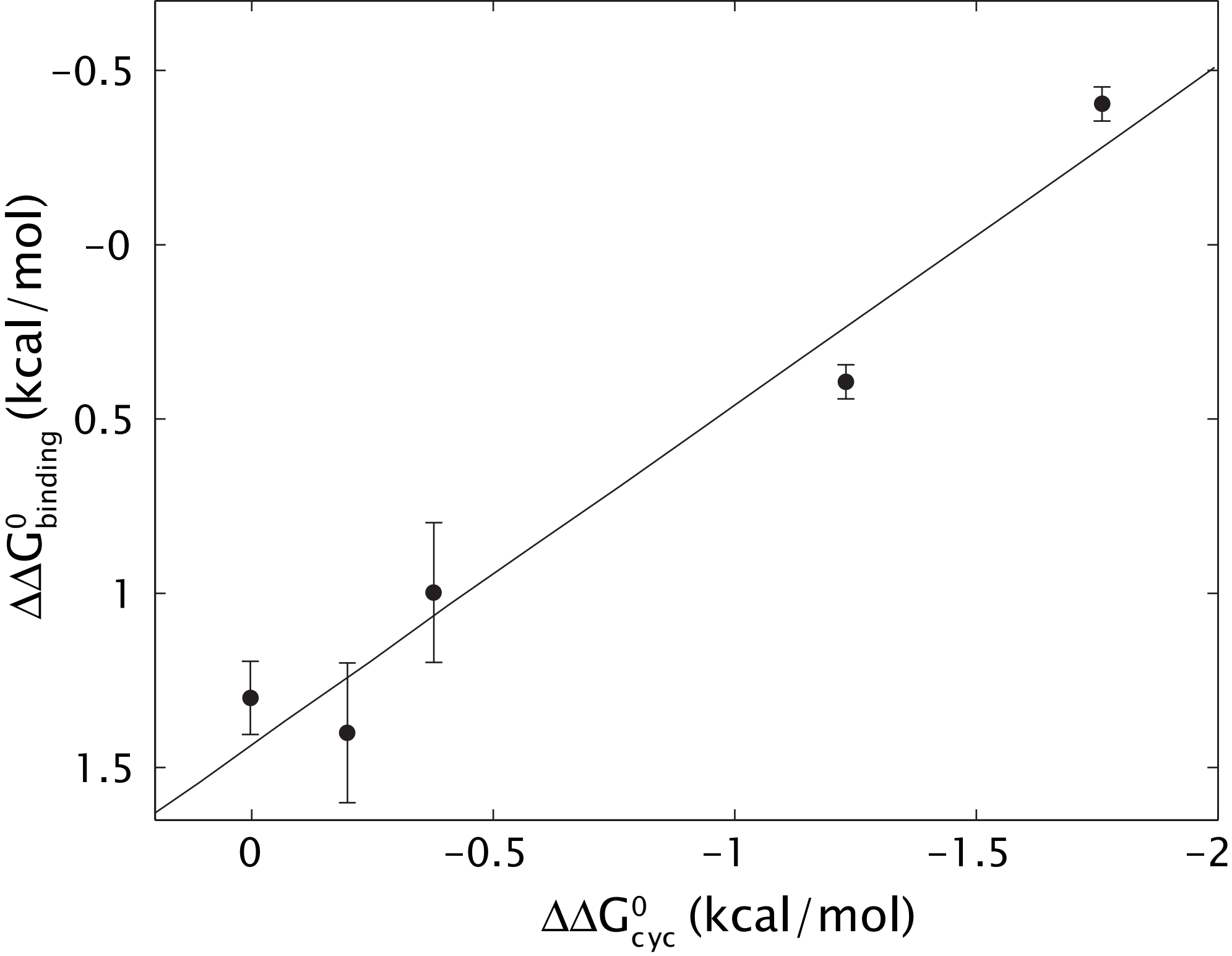

Figure 6

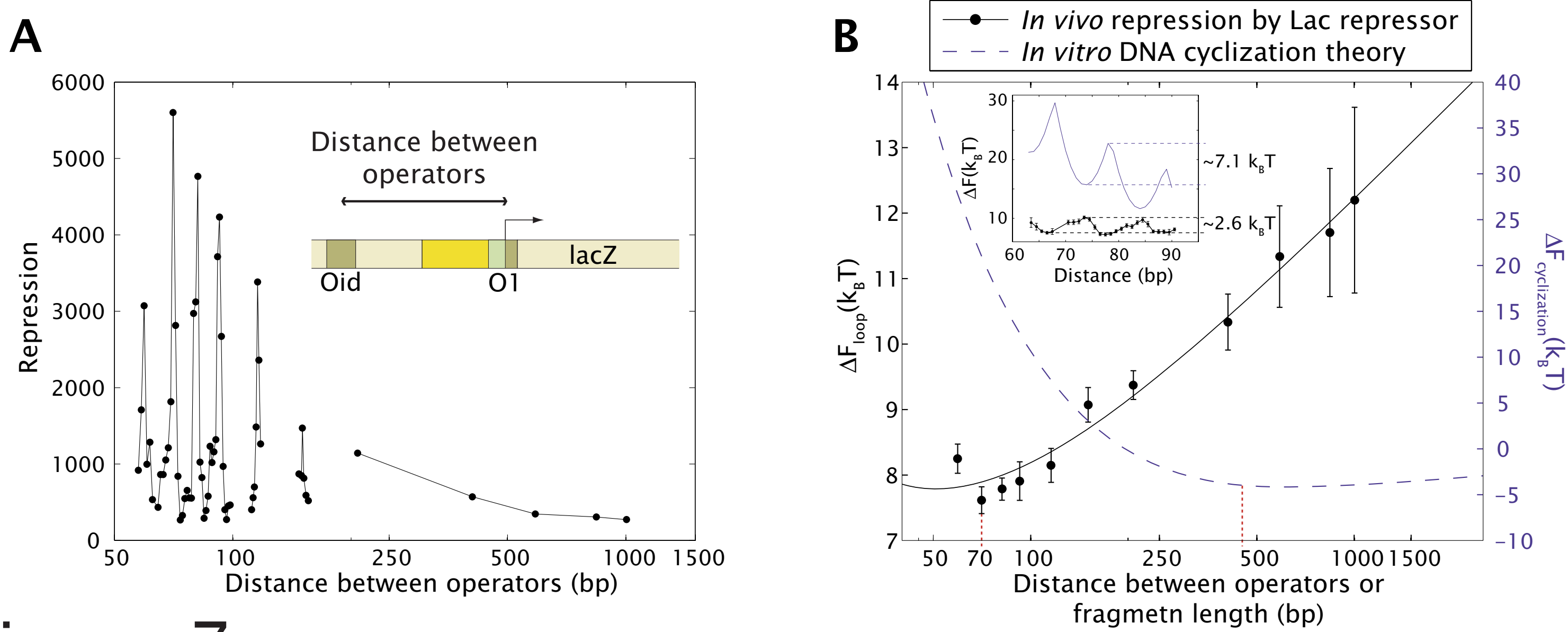
A
Distance between operators
Oid
O1
lacZ
Repression
Distance between operators (bp)
B
In vivo repression by Lac repressor
In vitro DNA cyclization theory
~7.1 kBT
~2.6 kBT
Distance (bp)
ΔF(kBT)
ΔFloop(kBT)
ΔFcyclization(kBT)
Distance between operators or fragmetn length (bp)

Figure 7

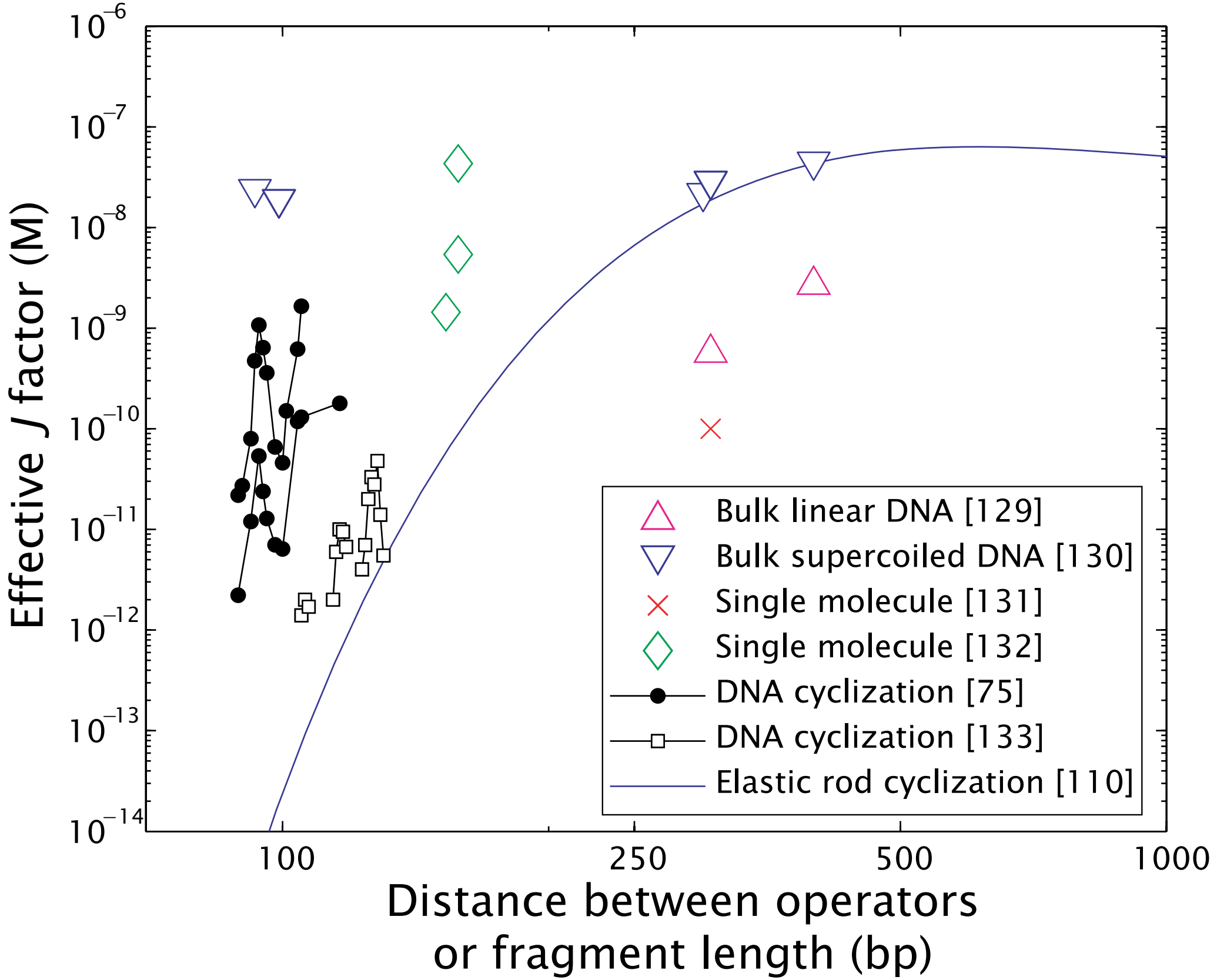

10^−6
10^−7
10^−8
10^−9
10^−10
10^−11
10^−12
10^−13
10^−14
Effective J factor (M)
100
250
500
1000
Distance between operators
or fragment length (bp)
Bulk linear DNA [129]
Bulk supercoiled DNA [130]
Single molecule [131]
Single molecule [132]
DNA cyclization [75]
DNA cyclization [133]
Elastic rod cyclization [110]

Figure 8

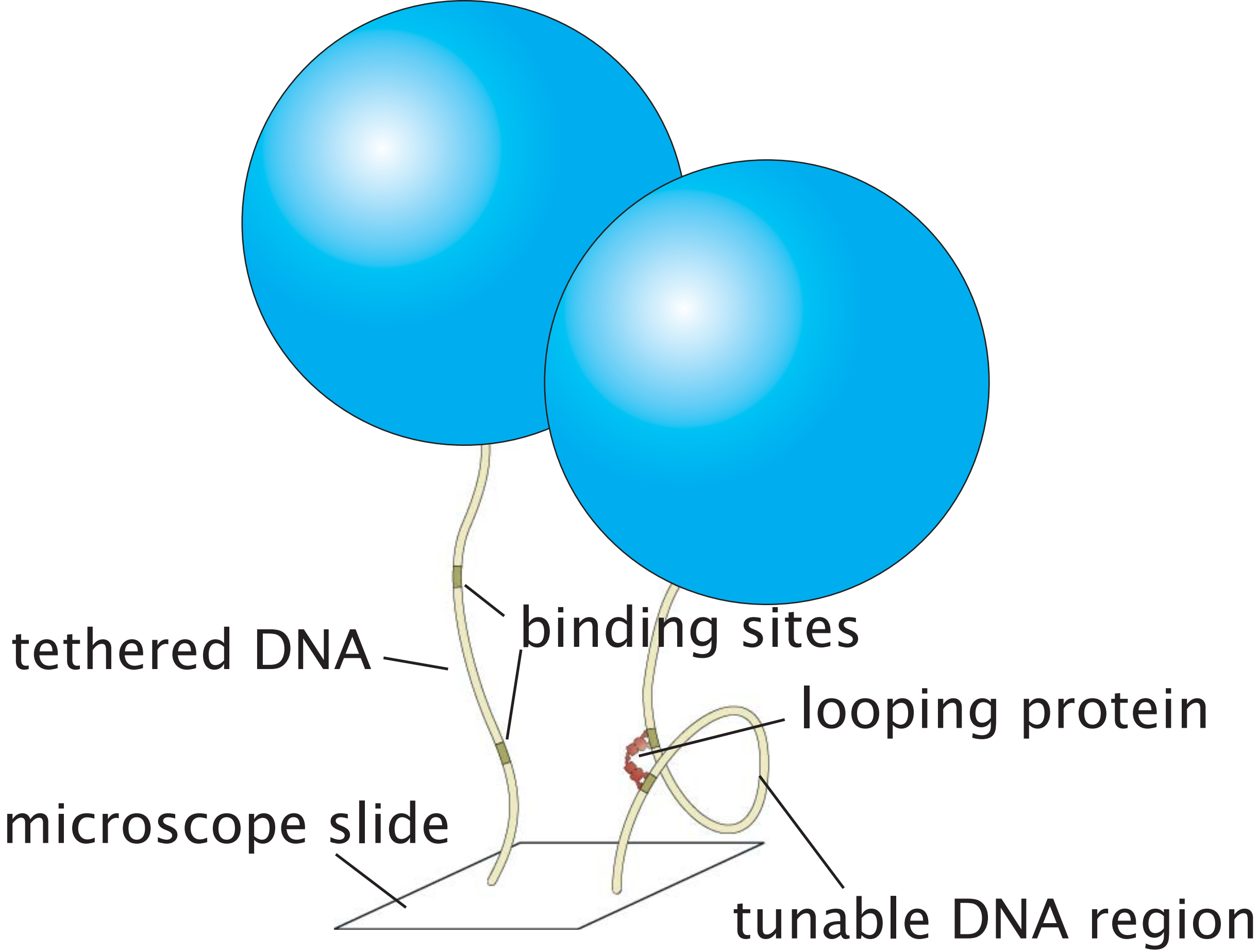
binding sites
tethered DNA
looping protein
microscope slide
tunable DNA region

Figure 9